\documentclass[conference,10pt]{IEEEtran}

\usepackage[T1]{fontenc}
\usepackage[utf8]{inputenc}
\usepackage{microtype}
\usepackage{graphicx}
\usepackage{booktabs}
\usepackage{array}
\usepackage{tabularx}
\usepackage{listings}
\usepackage{fancyvrb}
\usepackage{xcolor}
\usepackage{tikz}
\usetikzlibrary{positioning, arrows.meta, shapes.geometric, fit, calc,
                backgrounds, decorations.pathreplacing}
\usepackage{pgfplots}
\pgfplotsset{compat=1.18}
\usepackage{amsmath}
\usepackage{url}
\usepackage[hidelinks]{hyperref}
\usepackage{cleveref}

\definecolor{lstbg}{gray}{0.97}
\definecolor{lstkw}{rgb}{0.10,0.30,0.65}
\definecolor{lstcm}{rgb}{0.30,0.55,0.30}
\definecolor{lststr}{rgb}{0.65,0.10,0.10}

\lstdefinestyle{cstyle}{
  language=C,
  basicstyle=\ttfamily\footnotesize,
  keywordstyle=\color{lstkw}\bfseries,
  commentstyle=\color{lstcm}\itshape,
  stringstyle=\color{lststr},
  backgroundcolor=\color{lstbg},
  frame=single,
  rulecolor=\color{lstbg},
  breaklines=true,
  showstringspaces=false,
  columns=fullflexible,
  numbers=none,
  xleftmargin=0pt,
  xrightmargin=0pt,
  aboveskip=4pt,
  belowskip=4pt,
}
\lstdefinestyle{shellstyle}{
  basicstyle=\ttfamily\scriptsize,
  backgroundcolor=\color{lstbg},
  frame=single,
  rulecolor=\color{lstbg},
  breaklines=true,
  columns=fullflexible,
  numbers=none,
  xleftmargin=0pt,
  xrightmargin=0pt,
  aboveskip=4pt,
  belowskip=4pt,
}
\title{Cross-Core Inference Offload as an Operating-System Service\\
on Dual-Core Microcontrollers}

\author{\IEEEauthorblockN{Dimitrios Kafetzis}
\IEEEauthorblockA{SynapticOS Project, Hamburg, Germany}}

\begin{document}
\maketitle

\begin{abstract}
% Abstract — every numeric claim is anchored to Section 5 (Evaluation).
% Stub-NPU numbers are labeled as such; acceptance-criteria misses are
% reported as misses, per the honest-baseline methodology of Phases 1-2.
Commodity dual-core microcontrollers are asymmetric in a way their
marketing rarely admits: on NXP's MCXN947, the second Cortex-M33 has
no floating-point unit, no DSP extension, no TrustZone --- and no
memory protection unit. We argue this asymmetry is a design input,
not an obstacle, and present the Phase~3 dual-core architecture of
SynapticOS, an open-source AI-native runtime built on Zephyr: the
entire AI runtime (models, NPU and DSP access, the priority
scheduler) is confined to the capable core, and the application core
reaches inference exclusively through a message-based
operating-system service --- inference as a remote system call. The
transport is a pair of lock-free single-producer/single-consumer
rings in shared SRAM: every ring index has exactly one writer,
free-running 32-bit counters make wraparound safe by unsigned
arithmetic, and ordering needs only data-memory barriers --- no
cross-core atomics, which the platform does not offer. Because ring
state lives in the shared region rather than in either core's
private memory, an application core that reboots mid-conversation
rejoins the ring without runtime-core involvement. Requests carry
the scheduler's priority classes across the core boundary, errors
and timeouts propagate to the caller's return value, and tensors
stage zero-copy in a shared exchange slot --- a 27~KB camera frame
cannot afford to exist twice in the application core's 64~KB of RAM.
Measured on the FRDM-MCXN947 (both cores at 150~MHz): the
application core boots in 1{,}514~\textmu s from release and
completes the IPC handshake in 2{,}554~\textmu s, bit-identical
across 11 consecutive boots; message round-trip time is
15~\textmu s typical and 81~\textmu s worst-case against a
50~\textmu s budget; ring operations cost 25 cycles per push; a
1{,}913-serve alternating two-model soak completed with zero errors
(stub-NPU latencies, labeled --- these numbers bracket the runtime
and transport, not silicon inference throughput). A read-only MPU
region on the runtime core guards the application core's RAM ---
verified by an on-board fault-injection self-test --- and we state
the architectural limits honestly: the protection is one-directional
because the application core has no MPU, and ARMv8-M offers no way
to block privileged reads with the background map enabled. Two
defects that only the hardware could reveal are reported in full:
releasing the second core into erased flash wedges the entire chip
including the debug port (the fallback now blank-checks the flash
bank through the ROM API before release), and a Zephyr flash-driver
Kconfig silently disarmed the devicetree-declared MPU guard (the
guard is now programmed at runtime). The dual-core firmware fits in
98.9~KB of flash on the runtime core and 32.4~KB on the application
core (42.6~KB of its 64~KB RAM); a 108-test suite across 13 suites
passes 100\% under emulation. SynapticOS is released under
Apache~2.0 at
\url{https://github.com/Dimitrios-Kafetzis/SynapticOS}.

\end{abstract}

\begin{IEEEkeywords}
asymmetric multiprocessing, inter-processor communication, real-time
operating systems, lock-free data structures, memory protection,
neural processing units, edge AI, embedded systems
\end{IEEEkeywords}

\section{Introduction}
\label{sec:intro}

The first two papers in this project established an operating system
that treats neural inference as a first-class workload on a
single microcontroller core: Phase~1~\cite{synapticos-p1} built the
tensor-aware memory manager, the NPU hardware-abstraction layer, the
model registry, and the profiling surface; Phase~2~\cite{synapticos-p2}
promoted the inference pipeline itself to an OS object and put a
priority job scheduler in front of the accelerator. Both phases ran
entirely on one of the NXP MCXN947's two Cortex-M33 cores while the
other sat in reset --- which is how the overwhelming majority of
deployed dual-core MCU designs ship, because using the second core
means solving inter-processor communication, shared-memory layout,
boot orchestration, and cross-core fault containment before the
first application line runs.

This paper is about putting that second core to work --- and about
what the second core actually is. On the MCXN947, the two cores that
share a part number do not share a feature set: the CMSIS device
header for \texttt{cm33\_core1} sets \texttt{\_\_FPU\_PRESENT},
\texttt{\_\_DSP\_PRESENT}, \texttt{\_\_SAUREGION\_PRESENT}, and
\texttt{\_\_MPU\_PRESENT} all to zero. The second core cannot run
float-heavy preprocessing efficiently, cannot use the DSP
extension, sees the memory map through a different (non-secure)
address lens than its TrustZone-enabled sibling, and cannot protect
so much as a byte of its own RAM. A symmetric-multiprocessing
mindset treats each of these as a defect to work around. We take
the opposite position: the asymmetry \emph{is} the architecture.

\subsection{Inference as a Cross-Core Service}
\label{sec:intro:position}

Our design confines everything that needs the capable core to the
capable core: model storage and lifecycle, NPU and PowerQuad access,
the Phase~2 pipeline engine and priority scheduler, and the shell.
The application core --- the natural home for sensor polling,
protocol handling, and product logic --- gets none of that machinery
and needs none of it. What it gets instead is a service interface:
load a model by name, submit a tensor with a priority class, block
for the result with a timeout. The runtime core answers requests
from the application core through the same scheduler that serves its
own local jobs, so remote and local work contend in a single
priority space. In effect, inference becomes a remote system call
--- with the OS, not the application, owning the transport, the
staging memory, the priority semantics, and the failure modes.

Three properties of the platform shape the transport. First, there
is no cross-core lock or atomic worth trusting on this part, so the
message layer must need neither: two single-producer/%
single-consumer rings --- one per direction --- give every index
exactly one writer, and data-memory barriers are the only ordering
primitive used. Second, the application core has 64~KB of RAM and a
96$\times$96$\times$3 camera frame is 27~KB, so the request path
must not copy: tensors stage directly in a shared exchange slot.
Third, either core can reset while the other runs, so the ring
state itself lives in the shared region --- a rebooting
application core rejoins the conversation without the runtime core
doing anything at all.

\subsection{Contributions}
\label{sec:intro:contributions}

Building on the Phase~1 and Phase~2 subsystems behind the same
frozen public headers, this paper contributes:

\begin{enumerate}
\item An \textbf{asymmetric AMP architecture for MCU inference}
(\cref{sec:design}): the AI runtime confined to the capable core,
the application core accessing it exclusively through a
message-based service interface, with the capability asymmetry of
commodity dual-core silicon treated as a design input rather than
an obstacle.
\item A \textbf{lock-free SPSC message layer in shared SRAM}
(\cref{sec:design:ring}) requiring no cross-core lock or atomic
support: per-direction rings with single-writer indices,
free-running 32-bit counters (wraparound-safe by unsigned
arithmetic, tested at \texttt{UINT32\_MAX}), DMB-only ordering, and
64-byte index separation against false sharing. Measured cost:
25 cycles per push. Ring state resides in shared SRAM, yielding
reset resilience by construction.
\item \textbf{Inference-as-syscall semantics across cores}
(\cref{sec:design:infer}): requests carry the Phase~2 scheduler's
priority class so remote and local jobs contend in one priority
space; errors and timeouts propagate to the caller's return value;
input and output tensors stage zero-copy in a shared slot sized by
the application core's RAM constraint.
\item \textbf{One-directional hardware memory protection, honestly
scoped} (\cref{sec:design:mpu}): a read-only MPU region on the
runtime core guards the application core's RAM, with a fault policy
that aborts only the offending thread while both cores continue ---
verified by fault injection on the board. The architectural limits
are stated plainly: ARMv8-M offers no privileged-no-access encoding
while the background map is enabled (reads are not blockable), and
the application core has no MPU at all.
\item \textbf{Out-of-tree second-core enablement on mainline
Zephyr~3.7} (\cref{sec:impl}): a board port supplying the missing
\texttt{cpu1} target, the missing CPU-architecture Kconfig selects,
a corrected non-secure devicetree view, and per-core address
aliasing (secure $+$\texttt{0x1000\_0000} versus plain) folded into
one shared layout header that both images compile against, with
build-time asserts pinning every cross-core structure offset.
\item An \textbf{honest-baseline evaluation} (\cref{sec:eval})
continuing the Phase~1/2 methodology: all inference latencies run
through the deterministic stub NPU and are labeled as
runtime-and-transport baselines; the acceptance criteria the
measurements did not meet --- an NPU duty-cycle target predicated
on a saturation workload our demo deliberately is not, and two
negative paths not separately staged on the board --- are reported
as misses, not re-scoped. Two defects only the hardware could
reveal are documented in full (\cref{sec:discussion:defects}).
\end{enumerate}

\subsection{Scope}
\label{sec:intro:scope}

This paper reports the Phase~3 system as validated on the
\texttt{qemu\_cortex\_m3} continuous-integration target (SPSC ring
logic and shared-layout ABI, which are core-count-independent) and
live on the FRDM-MCXN947 board (all cross-core behaviour;
transcripts captured 2026-07-14, released as v0.3.0). The model
stage executes the deterministic stub NPU kernel inherited from
Phase~1: every latency that includes inference is a
runtime-and-transport baseline, not a silicon throughput claim. The
IPC timing numbers --- boot, handshake, round-trip, ring cost ---
measure real mechanisms end to end and are not stub-dependent.
One request is in flight at a time by design
(\cref{sec:design:infer}); the protection story is one-directional
by silicon (\cref{sec:design:mpu}); both limits are discussed
rather than hidden (\cref{sec:discussion:limits}).

\subsection{Paper Organisation}
\label{sec:intro:org}

\Cref{sec:background} lays out the dual-core reality of the MCXN947
and the gaps in mainline Zephyr~3.7 that had to be filled.
\Cref{sec:design} presents the shared-memory contract, the SPSC
rings, the boot protocol, the cross-core inference semantics, and
the protection model. \Cref{sec:impl} covers the board port and the
implementation of the zero-copy slot. \Cref{sec:eval} reports the
measurements; \cref{sec:related} positions the design against
OpenAMP/RPMsg, Zephyr's IPC services, and the lock-free queue
literature; \cref{sec:discussion} reconstructs the two board-found
defects and collects limitations. \Cref{sec:conclusion} concludes.

\section{Background: What a Dual-Core MCU Actually Is}
\label{sec:background}

\subsection{The MCXN947's Two Unequal Cores}
\label{sec:background:silicon}

The MCXN947~\cite{nxp-mcxn947} advertises two Cortex-M33 cores at
150~MHz. The datasheet's fine print, concentrated in the per-core
CMSIS device headers, is where the system architecture is actually
decided. CPU0 has a hardware FPU, the DSP extension, TrustZone-M
with SAU regions, and an 8-region MPU; it also owns the eIQ Neutron
NPU and PowerQuad DSP integration paths established in
Phases~1--2~\cite{synapticos-p1,synapticos-p2}. CPU1's header sets
\texttt{\_\_FPU\_PRESENT}, \texttt{\_\_DSP\_PRESENT},
\texttt{\_\_SAUREGION\_PRESENT}, and \texttt{\_\_MPU\_PRESENT} to
zero. Those four zeros reshaped three of this phase's five
deliverables before any code ran:

\begin{itemize}
\item \textbf{No FPU/DSP on CPU1} means float-heavy preprocessing
and the Phase~2 DSP paths must not migrate there --- confirming the
service architecture in which all tensor work stays on CPU0.
\item \textbf{No TrustZone on CPU1} means the two cores see the same
physical memory at different addresses: CPU0 runs Secure and uses
the $+$\texttt{0x1000\_0000} secure aliases (RAM at
\texttt{0x3xxx\_xxxx}, flash at \texttt{0x1xxx\_xxxx}), while CPU1
addresses RAM at \texttt{0x2xxx\_xxxx} and flash at
\texttt{0x0xxx\_xxxx}. Every shared pointer, every boot vector, and
every devicetree region must be expressed in the right lens for the
right core (\cref{sec:impl:alias}).
\item \textbf{No MPU on CPU1} means cross-core memory protection can
only ever be one-directional: CPU0 can be prevented from corrupting
CPU1's RAM, but nothing on this silicon can stop CPU1 from writing
anywhere it likes (\cref{sec:design:mpu}).
\end{itemize}

Memory is similarly less symmetric than the headline ``512~KB
SRAM'' suggests: 416~KB is contiguous main SRAM at
\texttt{0x2000\_0000} (banks RAMA--RAMH), while the remaining 96~KB
is RAMX on the code bus at \texttt{0x0400\_0000}. The phase plan's
original sketch gave CPU1 a 160~KB region ending at
\texttt{0x2007\_FFFF} --- an address that does not exist on the
main-SRAM bus. The map that ships (\cref{sec:design:layout}) is
256~KB for CPU0, 96~KB shared, and 64~KB for CPU1, which executes
in place from the second 1~MB flash bank at \texttt{0x0010\_0000}.
We flag this because plan-versus-datasheet corrections are exactly
the class of finding that only surfaces when a design is pushed all
the way to silicon, and reporting them is part of the methodology.

\subsection{What Mainline Zephyr 3.7 Does Not Provide}
\label{sec:background:zephyr}

Zephyr~3.7~\cite{zephyr} knows the MCXN947 exists --- CPU0 is a
supported build target --- but its second core is unreachable from
the mainline tree. Four gaps had to be filled out-of-tree, all
carried inside the SynapticOS repository so that a stock Zephyr
workspace builds both images:

\begin{enumerate}
\item \textbf{No CPU1 board target.} There is no
\texttt{frdm\_mcxn947/mcxn947/cpu1} board. We supply an out-of-tree
board port (\texttt{boards/nxp/frdm\_mcxn947\_cpu1}) with its own
devicetree, defconfig, and Kconfig, injected via
\texttt{BOARD\_ROOT}.
\item \textbf{The CPU1 SoC symbol selects no CPU architecture.}
\texttt{SOC\_MCXN947\_CPU1} exists in the tree but selects neither
\texttt{CPU\_CORTEX\_M33} nor its dependencies; the board Kconfig
merges the missing selects into the SoC symbol.
\item \textbf{The non-secure SoC devicetree include is incomplete.}
\texttt{nxp\_mcxn94x\_ns.dtsi} cannot be included stand-alone (it
lacks the memory and ARMv8-M includes its secure sibling gets
elsewhere), and it pulls in an Ethernet node whose binding demands
pinctrl the CPU1 image does not configure; the board devicetree
supplies the missing includes and deletes the offending node.
\item \textbf{The mailbox driver does not know this SoC.} Zephyr's
\texttt{mbox} driver lacks the MCXN947's CPU identifiers, so the
inter-core interrupt path uses NXP's header-only
\texttt{fsl\_mailbox.h} HAL directly (MAILBOX IRQ~54 on both
cores), the same pattern Phase~2 used for PowerQuad.
\end{enumerate}

None of these is deep; all of them are load-bearing, and each is a
small, upstreamable artifact. Their absence explains, in part, why
the second core of this popular part so often ships dark.

\subsection{Inherited Subsystems and the Frozen API}
\label{sec:background:inherited}

Phase~3 adds no new public header and modifies none: the IPC
surface (\texttt{syn\_ipc.h}) has been part of the frozen API since
Phase~1, when it was specified ahead of implementation ---
\texttt{syn\_ipc\_init()}, \texttt{send()}, \texttt{receive()} with
timeout, and per-type handler registration over a fixed 20-byte
message carrying an ID, a type, a priority, a payload reference,
a timestamp, and a status. This phase implements that
interface without amendment, which we take as evidence for the
Phase~1 bet that the IPC abstraction could be frozen before the
transport existed. The Phase~2 pipeline engine
and priority scheduler run unmodified on CPU0; cross-core requests
enter them through the same submission path local callers use.

\section{Design}
\label{sec:design}

\subsection{The Shared-Memory Contract}
\label{sec:design:layout}

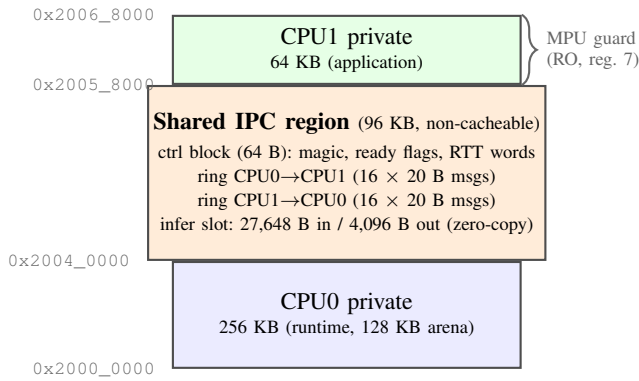
\begin{figure}[t]
\centering
\begin{tikzpicture}[
  font=\small,
  region/.style={rectangle, draw=black!70, thick,
    minimum width=46mm, align=center, inner sep=2pt},
  addr/.style={font=\scriptsize\ttfamily, text=black!60, anchor=east},
]

% --- main SRAM column (bottom = low address) ---
\node[region, minimum height=14mm, fill=blue!8]  (cpu0)
  at (0,0) {CPU0 private\\[-1pt]{\scriptsize 256 KB (runtime, 128 KB arena)}};
\node[region, minimum height=23mm, fill=orange!15, above=0mm of cpu0.north] (shared)
  {\textbf{Shared IPC region} {\scriptsize (96 KB, non-cacheable)}\\[1pt]
   {\scriptsize ctrl block (64 B): magic, ready flags, RTT words}\\[-1pt]
   {\scriptsize ring CPU0$\rightarrow$CPU1 (16 $\times$ 20 B msgs)}\\[-1pt]
   {\scriptsize ring CPU1$\rightarrow$CPU0 (16 $\times$ 20 B msgs)}\\[-1pt]
   {\scriptsize infer slot: 27{,}648 B in / 4{,}096 B out (zero-copy)}};
\node[region, minimum height=9mm, fill=green!10, above=0mm of shared.north] (cpu1)
  {CPU1 private\\[-1pt]{\scriptsize 64 KB (application)}};

% --- addresses on the left (CPU1 plain lens) ---
\node[addr] at ($(cpu0.south west)+(-1mm,0)$)  {0x2000\_0000};
\node[addr] at ($(shared.south west)+(-1mm,0)$){0x2004\_0000};
\node[addr] at ($(cpu1.south west)+(-1mm,0)$)  {0x2005\_8000};
\node[addr] at ($(cpu1.north west)+(-1mm,0)$)  {0x2006\_8000};

% --- MPU guard marker on CPU1 region ---
\draw[decorate, decoration={brace, amplitude=2mm}, thick, black!60]
  (cpu1.north east) -- (cpu1.south east)
  node[midway, right=2mm, font=\scriptsize, align=left, text=black!60]
  {MPU guard\\(RO, reg.\ 7)};

\end{tikzpicture}
\caption{The 416~KB main SRAM tiled between the cores, and the
shared region's internals (CPU1's plain address lens shown; CPU0,
running Secure, sees the same bytes at $+$\texttt{0x1000\_0000}).
Offsets, sizes, and field layouts are pinned by
\texttt{BUILD\_ASSERT}s in the one header both images compile;
payloads are addressed by offset because the two cores see the same
bytes through different address lenses. CPU0's last MPU region
guards CPU1's RAM read-only (\cref{sec:design:mpu}).}
\label{fig:memmap}
\end{figure}

Everything the two cores agree on is written down in one private
header, \texttt{syn\_shared\_layout.h}, which both images compile.
\Cref{fig:memmap} shows the resulting map: the 416~KB of main SRAM
tiles into 256~KB of CPU0-private memory, a 96~KB shared IPC region,
and 64~KB of CPU1-private memory, with no gaps and no overlap ---
and the header carries \texttt{BUILD\_ASSERT}s proving exactly that,
plus asserts pinning the size and every cross-core field offset of
each shared structure. A layout mismatch between images is a compile
error, not a Heisenbug: the 20-byte message, the 64-byte control
block, the ring geometry, and the exchange-slot header are all
pinned bit-for-bit.

% Figure 1 — SRAM map and shared-region internals, with per-core address lenses.

The shared region opens with a control block holding a magic value
(\texttt{"SYN3"}), a layout version, the ring geometry, two ready
flags used by the boot handshake (\cref{sec:design:boot}), and four
words CPU1 uses to publish round-trip statistics
(\cref{sec:eval:rtt}) --- a deliberate design for a core with no
console of its own. The two rings follow, then a payload pool
addressed by offset (never by pointer: an offset means the same
thing in both cores' address lenses; a pointer does not).

\subsection{Lock-Free SPSC Rings}
\label{sec:design:ring}

The transport must function without cross-core mutual exclusion,
because the platform offers none worth trusting: there is no shared
lock peripheral used here, and exclusive-monitor semantics across
two masters on this interconnect are exactly the kind of thing a
correctness argument should not lean on. The design therefore
guarantees by \emph{construction} that no location is ever written
by two cores.

\begin{lstlisting}[float=tp,caption={The shared ring (abridged from
\texttt{syn\_shared\_layout.h}). Only the producer core writes
\texttt{head}; only the consumer core writes \texttt{tail}; each
index sits alone in a 64-byte block.},label={lst:ring},captionpos=b]
typedef struct {
    volatile uint32_t head;  /* producer-written */
    uint32_t _pad0[15];
    volatile uint32_t tail;  /* consumer-written */
    uint32_t _pad1[15];
    syn_ipc_msg_t slots[SYN_IPC_RING_ENTRIES];
} syn_ipc_ring_t;
\end{lstlisting}

Two rings, one per direction, give each core a single fixed role
per ring: CPU0 produces into \texttt{ring\_c0\_to\_c1} and consumes
from \texttt{ring\_c1\_to\_c0}; CPU1 the reverse. Within a ring
(\cref{lst:ring}), the indices are free-running 32-bit counters:
slot $=$ \texttt{head \% N}, the ring is full when
\texttt{head $-$ tail $=$ N}, and unsigned wraparound at
$2^{32}$ is handled by the arithmetic itself --- a unit test drives
the counters across \texttt{UINT32\_MAX} to hold the claim. This is
Lamport's classic result~\cite{lamport77} operationalised: a
single-producer/single-consumer queue needs no lock, only ordered
visibility. On ARMv8-M~\cite{armv8m} that ordering is two
\texttt{DMB} barriers: the producer writes the slot, barriers, then
publishes \texttt{head}; the consumer reads \texttt{head}, barriers,
then reads the slot. The 64-byte separation between \texttt{head}
and \texttt{tail} is the false-sharing discipline of the
high-performance queue literature~\cite{fastforward,disruptor}
applied at MCU scale --- cheap insurance on today's
non-coherent-cache parts that becomes load-bearing the moment a
cached sibling of this design exists.

The measured cost is 25 cycles per push and 41 cycles per
push-plus-pop pair including the barriers
(\cref{sec:eval:ringcost}): at 150~MHz, ring mechanics contribute
about 167~ns to a message hop. Everything else in the round-trip
budget is interrupt latency and thread wakeup, which is where it
belongs.

One consequence of keeping ring state in shared SRAM deserves
promotion to a design principle: \textbf{the ring is the durable
party in the conversation}. Neither core's private state is needed
to resume messaging, so when the application core reboots ---
watchdog, brownout, deliberate reset --- it re-attaches to live
indices and the conversation continues. \Cref{sec:eval:reset}
exercises this eleven times on the board.

Above the rings, a thin dispatch layer preserves the frozen API's
two consumption styles without breaking the single-consumer
invariant: one dispatch thread per core is the sole ring consumer,
woken by the MAILBOX inter-core interrupt; registered per-type
handlers run on it, and message types without a handler are
forwarded to an internal queue that \texttt{syn\_ipc\_receive()}
blocks on with a timeout. Handlers therefore execute in thread
context (not ISR context), and one slow handler delays only its own
core's dispatch --- a documented trade
(\cref{sec:discussion:limits}).

\subsection{Boot Protocol}
\label{sec:design:boot}

CPU0 owns the boot sequence end to end. After the runtime
initialises, it (1)~zeroes and stamps the shared control block,
(2)~sets \texttt{cpu0\_ready}, (3)~\emph{verifies that flash
bank~1 actually contains an image}, (4)~writes CPU1's vector-table
address into SYSCON's \texttt{CPBOOT} --- expressed in CPU1's plain
address lens, \texttt{0x0010\_0000} --- and releases the core via
\texttt{CPUCTRL}, then (5)~polls \texttt{cpu1\_ready} with a
timeout. CPU1's image, on booting, attaches to the shared region,
validates magic, version, and ring geometry, sets its ready flag,
and sends a \texttt{STATUS\_REQ} as a first-light handshake message.
On the board, release-to-ready is 1{,}514~\textmu s and
release-to-handshake 2{,}554~\textmu s
(\cref{sec:eval:boot}) --- against budgets of 100 and 200~ms.

Step~(3) is not defensive boilerplate; it is the scar tissue of the
most instructive failure of the phase. The original fallback design
--- release CPU1, wait for the handshake, time out, log, continue
single-core --- is unsurvivable on this silicon: reads of erased
flash raise ECC bus errors, and a CPU1 vector fetch from a blank
bank stalls the flash controller that CPU0 is executing in place
from. The observable result is a chip with no serial output and a
dead debug port (\cref{sec:discussion:defects}). The shipped
fallback therefore interrogates bank~1 through the MCX ROM API's
flash-controller commands (\texttt{FLASH\_VerifyErase} plus a
\texttt{FLASH\_Read} and a vector sanity check) --- commands that go
through the flash management controller rather than the bus, and so
are safe against erased pages --- \emph{before} touching
\texttt{CPBOOT}. A blank bank logs
\texttt{CPU1 image absent: single-core mode} and the full
single-core system comes up, shell included
(\cref{sec:eval:fallback}).

\subsection{Inference as a Remote System Call}
\label{sec:design:infer}

The application core's inference interface is deliberately shaped
like a blocking system call, because that is the semantics
application code wants: name a model, hand over a tensor and a
priority, block with a timeout, get a return value that is either a
result length or a negative errno.

Under the hood, four of the frozen message types carry the
protocol. \texttt{MODEL\_LOAD} resolves a model name to a handle on
the runtime core (the model itself already lives there; nothing
crosses but the name and the handle). \texttt{INFER\_REQ} stakes
the request: the caller's tensor is already staged in the shared
exchange slot at that point, written there \emph{directly} by the
application --- the slot is the working buffer, not a copy target.
The 27{,}648-byte camera frame that motivates this cannot exist
twice on a 64~KB core; zero-copy staging is not an optimisation
here but an admission requirement, and it removes a memcpy from the
hot path as a side effect. On the runtime core, the serving layer
resolves the handle, submits the tensor to the Phase~2 scheduler
\emph{with the priority class carried by the message} --- remote
\textsc{realtime} outranks local \textsc{normal}, because there is
one priority space, not one per core --- and answers with
\texttt{INFER\_RESP}, whose status field carries the inference
result code. Timeouts surface as \texttt{-ETIMEDOUT} from the
blocking call; a runtime-side failure rides back in the response
status. The error path is implemented and unit-tested, though no
organic NPU failure occurred on the board to exercise it end to end
--- reported as such in \cref{sec:eval:accept}.

The slot holds one request at a time; the application-core helper
serialises callers with a local mutex, so concurrent requesters
queue behind the slot rather than interleaving in the ring. This is
a documented capacity decision, not an accident
(\cref{sec:discussion:limits}): the target workloads are
sensor-cadence request streams, the measured service time is
milliseconds, and a second in-flight slot buys nothing until the
NPU itself is the bottleneck.

\subsection{The Protection Model, Honestly Scoped}
\label{sec:design:mpu}

What can hardware actually enforce between these two cores? The
honest answer --- less than the phase plan hoped, for architectural
reasons worth publishing --- comes in three parts.

\textbf{CPU0 writes to CPU1's RAM: blockable, and blocked.} The
runtime core programs its final MPU region (region~7) over CPU1's
64~KB at boot, read-only with an explicit any-privilege-level
read-only access encoding. A write from any CPU0 thread into CPU1's
memory raises a MemManage fault with the offending address in
MMFAR. The fault policy is containment, not panic: an overridden
fatal-error handler logs the full dump and aborts only the
offending thread; the shell, the runtime, and cross-core traffic
all continue. \Cref{sec:eval:mpu} shows the self-test doing exactly
this on the board while inference serving runs uninterrupted.

\textbf{CPU0 reads of CPU1's RAM: not blockable.} ARMv8-M's MPU has
no privileged-no-access encoding when the privileged background map
is enabled~\cite{armv8m}, and disabling the background map
(\texttt{PRIVDEFENA=0}) on a core that must run ROM API calls and
touch peripherals behind Zephyr's back is a reliability trade we
rejected. Confidentiality between the cores is therefore not
claimed --- only integrity, in one direction.

\textbf{CPU1 accesses to CPU0's RAM: not constrainable at all.}
CPU1 has no MPU. Nothing on this silicon can stop the application
core from writing anywhere in the map. The protection story is
one-directional \emph{by construction of the part}, and we consider
saying so plainly more useful than the common alternative of not
mentioning it: on asymmetric silicon, the correct threat model
protects the deterministic, multi-tenant runtime core from the
application core's likely bugs where possible --- and where the
hardware cannot express that, the design must rely on the service
interface being the only surface the application is given.

One more property of this subsystem was board-taught rather than
designed, and became a second design rule: \textbf{the guard is
programmed at runtime, not declared in devicetree}, because a
devicetree-declared read-only attribute turned out to be silently
rewritable by an unrelated Kconfig select
(\cref{sec:discussion:defects}). The runtime programming happens in
a \texttt{SYS\_INIT} hook ordered after the architecture MPU
initialisation, and a shell-invocable self-test (\texttt{syn mpu
test}) re-verifies the guard on demand --- which is how the
regression was caught.

\section{Implementation}
\label{sec:impl}

Phase~3 adds roughly 2{,}500 lines to the repository (2{,}547
insertions since v0.2.0, tests included) behind the frozen public
headers: the ring core and dispatch layer, the boot orchestrator,
the MPU guard and fault policy, the cross-core serving layer, the
out-of-tree CPU1 board port, and the \texttt{dual\_model}
demonstration pair. This section covers the three implementation
problems whose solutions are least visible in the design story.

\subsection{Per-Core Address Aliasing in One Header}
\label{sec:impl:alias}

Because CPU0 runs Secure and CPU1 has no TrustZone, the same
physical byte has two addresses --- and both images must agree on
\emph{physical} layout while each dereferences pointers in its own
lens. The shared layout header folds the difference into a single
constant: \texttt{SYN\_SHM\_ALIAS} is \texttt{0x1000\_0000} when
compiling for CPU0 and zero for CPU1, and every base address in the
map is expressed as \emph{plain address $+$ alias}. Code on either
side simply uses \texttt{SYN\_SHM\_SHARED\_BASE} and gets a pointer
it can dereference. Two places must escape the fold, and both are
annotated loudly in the source: the \texttt{CPBOOT} vector CPU0
writes for CPU1 must be in CPU1's plain lens (CPU1 fetches its own
vectors), and payload references inside messages are offsets from
the shared-region base rather than pointers, so they are
lens-independent by construction.

\subsection{Measuring Round-Trip Time from a Core with No Console}
\label{sec:impl:rtt}

The FRDM board's virtual COM port belongs to CPU0, and giving CPU1
a console would distort exactly the code being measured. The
round-trip measurement therefore runs where the latency is
experienced --- on CPU1 --- and publishes where the observer is:
CPU1 timestamps a \texttt{STATUS\_REQ} send with its cycle counter,
computes the delta in its \texttt{STATUS\_RESP} handler (on the
dispatch thread, so the number includes the MAILBOX interrupt, the
wakeup, and dispatch --- the full price an application pays), and
folds last/min/max/count into the four spare words of the shared
control block. CPU0's \texttt{syn ipc status} shell command reads
them out. The methodology costs four words of shared RAM and no
instrumentation on the measured path beyond one cycle-counter read
at each end.

\subsection{The Serving Layer}
\label{sec:impl:serving}

On CPU0, cross-core serving is a registered message handler, not a
privileged subsystem: \texttt{MODEL\_LOAD} performs a registry
lookup by name and answers with the handle;
\texttt{INFER\_REQ} wraps the staged input in a tensor descriptor
pointing \emph{into the shared slot} (zero-copy on the serving side
too), submits it to the Phase~2 scheduler with the message's
priority class, copies the result into the slot's output area, and
answers \texttt{INFER\_RESP} with the scheduler's status code.
Serve statistics (count, errors, average latency) accumulate for
the shell. Since handlers run on the dispatch thread, a
long-running inference delays heartbeat handling on the same core
--- measurably: it is the reason the observed worst-case RTT rises
from 15 to 81~\textmu s when a request lands mid-inference
(\cref{sec:eval:rtt}).

\subsection{Two Images, One Library}
\label{sec:impl:build}

Both images link the same \texttt{synaptic\_os} library; the CPU1
build differs only in configuration (no shell, no PowerQuad, no
boot orchestrator --- guarded by the CPU0-only Kconfig) and in its
board target. The CPU1 application is under 200 lines: attach, load
two models by name, alternate inference requests at sensor cadence,
publish RTT. Everything else --- ring discipline, dispatch,
staging, timeout handling --- is library code identical on both
cores, which is what makes the single-core QEMU test suite
meaningful for cross-core correctness: the SPSC logic under test is
the very object file the board runs.

\section{Evaluation}
\label{sec:eval}

We evaluate with the methodology of the previous two
papers~\cite{synapticos-p1,synapticos-p2}: numbers that pass
through the deterministic stub NPU are labeled
runtime-and-transport baselines rather than silicon inference
throughput; the IPC mechanism numbers (boot, handshake, ring cost,
round-trip) measure real hardware end to end and carry no stub
caveat; and acceptance criteria the measurements did not meet are
reported as misses. QEMU results were captured on 2026-07-14
(\texttt{community/phase3/results-qemu.md}); FRDM-MCXN947 results
were captured live on the board the same day from the v0.3.0
release-candidate build, with the raw serial transcripts in the
repository (\texttt{community/phase3/serial-frdm-*.log}) alongside
the consolidated \texttt{results-frdm.md}.

\subsection{Experimental Setup}
\label{sec:eval:setup}

The board runs the \texttt{dual\_model} sample: CPU0 boots the full
runtime (128~KB arena, shell, PowerQuad, cross-core serving),
registers a face-detection and a keyword-spotting model (stub
blobs), releases CPU1, and serves; CPU1 runs the remote image on
the out-of-tree board target, alternating \textsc{realtime}
face-detection and \textsc{normal} keyword-spotting requests at a
20~Hz sensor cadence. Both cores run at 150~MHz on Zephyr
v3.7.0~\cite{zephyr} under SDK~0.16.8 with \texttt{-Os}. QEMU
(\texttt{qemu\_cortex\_m3}) is single-core: it runs the 108-case
unit suite, exercising the SPSC ring logic with both roles driven
from one core --- the identical object code the board runs
cross-core (\cref{sec:impl:build}) --- plus the shared-layout ABI
asserts and the 10{,}000-message integrity sweep.

\subsection{Boot and Handshake}
\label{sec:eval:boot}

From CPU0 writing \texttt{CPUCTRL} to CPU1 setting its ready flag:
\textbf{1{,}514~\textmu s}, a 66$\times$ margin against the 100~ms
acceptance budget. From release to the first handshake message
completing: \textbf{2{,}554--2{,}577~\textmu s} against a 200~ms
budget. Across a deliberate series of eleven consecutive
power-on/reset cycles, every boot logged \emph{bit-identical}
timing (``CPU1 boot 1514 us, handshake 2577 us'' eleven times in
the transcript) --- the dual-core bring-up is not just fast but
deterministic, which matters more than the margin: a boot path with
variance would be hiding a race.

\subsection{Ring Operation Cost}
\label{sec:eval:ringcost}

The QEMU cycle-counted benchmark over 10{,}000 round-trips through
the real ring code measures \textbf{25 cycles per push}, 25 per
pop, and 41 per push-plus-pop pair, barriers included (QEMU
executes \texttt{DMB} as a no-op, but every load and store on the
path is real; the board-side barrier cost is inside the round-trip
number below). At 150~MHz, ring mechanics are $\sim$167~ns per
message hop: the transport's own cost is two orders of magnitude
below the round-trip budget, exactly where a message layer should
sit.

\subsection{Round-Trip Time}
\label{sec:eval:rtt}

Measured by CPU1 itself (\cref{sec:impl:rtt}) over 64
heartbeat round-trips during live two-model serving:
\textbf{last 15~\textmu s, minimum 15~\textmu s, maximum
81~\textmu s} against the 50~\textmu s acceptance budget. The
number includes the MAILBOX interrupt, dispatch-thread wakeup, and
handler dispatch on both cores --- the full application-visible
price. The typical case beats the budget by 3.3$\times$; the
worst case exceeds it, and we report it rather than filtering it:
the 81~\textmu s tail occurs when a heartbeat lands while CPU0's
dispatch thread is mid-inference (\cref{sec:impl:serving}), a
direct, explainable consequence of running handlers on the
dispatch thread. \Cref{lst:ipcstatus} shows the shell view,
verbatim from the transcript.

\begin{lstlisting}[style=shellstyle,float=tp,caption={\texttt{syn ipc
status} on the FRDM-MCXN947 during the release-candidate soak
(verbatim from \texttt{serial-frdm-final.log}).},
label={lst:ipcstatus},captionpos=b]
uart:~$ syn ipc status
CPU1 link: UP
CPU1 boot time: 1514 us (release to ready)
IPC handshake:  2554 us (release to STATUS_REQ)
STATUS_REQ answered: 64
Inferences served: 1267 (errors 0, avg 2290 us)
IPC round-trip (CPU1-measured, 64 samples):
  last 15 us, min 15 us, max 81 us
\end{lstlisting}

\subsection{Cross-Core Inference and Soak}
\label{sec:eval:infer}

End to end --- CPU1 stages a 27{,}648-byte 96$\times$96$\times$3
frame, sends \texttt{INFER\_REQ}, CPU0 schedules and runs the
(stub) model, answers, CPU1 wakes with the result --- the first
face-detection serve completes in \textbf{3{,}470~\textmu s}, and
the running average across both alternating models settles at
\textbf{2{,}290~\textmu s} (stub-NPU baseline, labeled; the
acceptance budget, set for real workloads, is 150~ms). The
alternating soak ran to \textbf{1{,}913 serves with zero errors} in
its longest single session --- roughly 4{,}000 ring messages
including heartbeats --- with earlier sessions recording 2{,}501+
and the final-image capture 1{,}267, all error-free. The QEMU
integrity sweep complements the board soak: 10{,}000 messages in
which every field of every message is derived from its sequence
number and checked on pop, across bursts sized to hit every ring
fill level, with zero loss and zero corruption.

\subsection{Memory Protection Under Fire}
\label{sec:eval:mpu}

\Cref{lst:mputest} reproduces the on-board self-test verbatim: a
shared-region write/readback succeeds, then a deliberate write of
\texttt{0xdeadbeef} into CPU1's RAM raises a MemManage fault at the
guarded address (MMFAR \texttt{0x3006\_0000}); the fault policy
logs the full dump, aborts the offending thread, and the transcript
shows serving traffic continuing across the fault --- CPU1 never
noticed. This is the containment contract of \cref{sec:design:mpu}
demonstrated live, on the runtime-programmed region~7 guard that
replaced the Kconfig-vulnerable devicetree version
(\cref{sec:discussion:defects}).

\begin{lstlisting}[style=shellstyle,float=tp,caption={The MPU
self-test on the board (abridged: interleaved serving-traffic log
lines elided). The fault dump is expected output; the offending
thread is aborted and both cores continue.},
label={lst:mputest},captionpos=b]
uart:~$ syn mpu test
Running cross-core MPU self-test (a MemManage
fault dump below is EXPECTED)...
MPU self-test PASS: shared region writable,
cross-core write faulted
<inf> syn_mpu: Shared region write/readback OK
<err> os: ***** MPU FAULT *****
<err> os:   Data Access Violation
<err> os:   MMFAR Address: 0x30060000
<err> os: r2/a3:  0xdeadbeef
<err> os: >>> ZEPHYR FATAL ERROR 19
<err> syn_mpu: MPU violation (reason 19):
  aborting offending thread, core continues
<inf> syn_mpu: Cross-core write to 0x30060000
  faulted as expected
\end{lstlisting}

\subsection{Single-Core Fallback}
\label{sec:eval:fallback}

With flash bank~1 deliberately erased, the release-candidate image
boots single-core: the ROM-API blank check detects the missing
image \emph{before} any release, logs \texttt{CPU1 image absent
(flash bank 1 blank): single-core mode}, and the full CPU0 system
comes up with the shell live and \texttt{syn ipc status} reporting
\texttt{CPU1 link: DOWN}. The reason this path exists in its
present form --- the original release-then-timeout design wedged
the entire chip --- is reconstructed in
\cref{sec:discussion:defects}.

\subsection{Reset Resilience}
\label{sec:eval:reset}

Ring indices live in shared SRAM (\cref{sec:design:ring}), so a
rebooting CPU1 rejoins without CPU0's involvement. The eleven
whole-chip reset cycles of \cref{sec:eval:boot} exercise the full
rejoin path end to end --- identical timing, clean handshake, and
resumed serving every time. An \emph{independent} CPU1-only reset
mid-conversation cannot be triggered from the board's buttons and
was not separately staged; we report the claim at the strength the
evidence supports.

\subsection{Footprint}
\label{sec:eval:size}

\Cref{tab:size3} reports the release-candidate images. The
application-core image --- the complete remote client including
the IPC layer and both model bindings --- is 32.4~KB of flash and
fits its working set in 42.6~KB of the core's 64~KB RAM. The 64~KB
constraint earned its keep during development: the first linked
remote image overflowed RAM by 4.7~KB with a private frame buffer,
which is precisely what forced the zero-copy staging design
(\cref{sec:design:infer}) --- the constraint improved the
architecture.

\begin{table}[t]
\caption{Release-candidate build footprints (flash = text+data,
RAM = data+bss).}
\label{tab:size3}
\centering
\renewcommand{\arraystretch}{1.15}
\setlength{\tabcolsep}{4pt}
\begin{tabular}{@{}llrr@{}}
\toprule
\textbf{Image} & \textbf{Target} & \textbf{Flash} & \textbf{RAM} \\
\midrule
dual\_model CPU0 (runtime, shell, serve) & FRDM & 98.9 KB & 205.9 KB \\
dual\_model CPU1 (remote client)         & FRDM & 32.4 KB & 42.6 KB \\
hello\_inference (regression)            & QEMU & 42.0 KB & 30.4 KB \\
\bottomrule
\end{tabular}
\end{table}

\subsection{Test Suite}
\label{sec:eval:tests}

The suite grows from Phase~2's 99 cases to \textbf{108 cases in 13
suites, 100\% pass} on \texttt{qemu\_cortex\_m3} via twister. The
growth is concentrated where the phase worked:
\texttt{syn\_ipc\_suite} goes from one placeholder case to ten real
ones --- 20-byte wire-format ABI pinning, shared-layout offset
pinning, memory-map tiling, empty/full \texttt{-EAGAIN} semantics,
FIFO order, the 10{,}000-message integrity sweep, index wraparound
at \texttt{UINT32\_MAX}, and the ring cost benchmark. No existing
test changed, consistent with no frozen header changing.

\subsection{Acceptance Criteria, Including the Misses}
\label{sec:eval:accept}

\Cref{tab:accept} consolidates the phase plan's acceptance criteria
against what was measured. Three rows deserve the honesty the
methodology demands. \textbf{NPU utilisation:} the plan's
``$>$60\%'' presumes a saturation workload; \texttt{dual\_model} is
deliberately cadence-paced at 20~Hz, giving $\sim$6\% NPU duty
cycle, while within each inference the NPU is busy 1{,}049 of
1{,}050~\textmu s (99.9\%, \texttt{syn prof last}). We report the
6\% as measured rather than re-staging the demo to flatter the
criterion. \textbf{Error propagation:} the path is implemented and
unit-tested, but no NPU failure occurred organically on the board,
so the negative path is not board-demonstrated.
\textbf{Priority preemption:} request priority is carried and fed
to the scheduler throughout (the alternating \textsc{realtime}/%
\textsc{normal} soak), but a staged contention experiment ---
remote \textsc{realtime} arriving mid-flight against a running
local \textsc{normal} job --- was not separately run; the
scheduler's priority semantics are Phase~2-verified.

\begin{table}[t]
\caption{Phase 3 acceptance criteria versus measurement (FRDM
unless noted). Stub-NPU rows are runtime-and-transport baselines.}
\label{tab:accept}
\centering
\footnotesize
\renewcommand{\arraystretch}{1.25}
\setlength{\tabcolsep}{3pt}
\begin{tabularx}{\columnwidth}{@{}lX@{}}
\toprule
\textbf{Criterion (budget)} & \textbf{Measured} \\
\midrule
CPU1 boot ($<$100 ms) & 1{,}514 \textmu s; identical over 11 boots \\
Handshake ($<$200 ms) & 2{,}554--2{,}577 \textmu s \\
RTT ($<$50 \textmu s) & 15 \textmu s typical; 81 \textmu s worst case
  (mid-inference arrival), reported as measured \\
10k msgs, no loss & QEMU sweep: 0 loss / 0 corruption; board soak:
  $\sim$4{,}000 msgs, 0 errors \\
Full/empty semantics & \texttt{-EAGAIN} / timeout paths, QEMU-verified \\
E2E inference ($<$150 ms) & 3{,}470 \textmu s first serve; 2{,}290
  \textmu s avg (stub NPU, labeled) \\
1{,}000-cycle soak & 1{,}913 serves, 0 errors (longest session) \\
Cross-core write faults & PASS: fault + thread abort + both cores
  continue (\cref{lst:mputest}) \\
CPU1$\to$CPU0 protection & NOT POSSIBLE: CPU1 has no MPU (silicon) \\
CPU0 reads blocked & NOT POSSIBLE: no ARMv8-M encoding with
  background map on \\
CPU1-absent fallback & PASS: ROM-API blank check, single-core boot \\
NPU utilisation ($>$60\%) & NOT MET as duty cycle: $\sim$6\%
  (20 Hz cadence-paced demo); 99.9\% within each inference \\
Error propagation & Implemented + unit-tested; not
  board-demonstrated (no organic failure) \\
Priority across cores & Carried and scheduled throughout; staged
  preemption experiment not run \\
\bottomrule
\end{tabularx}
\end{table}

\section{Related Work}
\label{sec:related}

The Phase~1 and Phase~2 papers surveyed the MCU inference-runtime
landscape and the pipeline/scheduler
literature~\cite{synapticos-p1,synapticos-p2}; this section focuses
on the questions Phase~3 answers: how cores talk, how the
conversation is protected, and where inference offload sits in the
taxonomy of heterogeneous execution.

\subsection{AMP Messaging Stacks}
\label{sec:related:amp}

OpenAMP~\cite{openamp} with RPMsg over virtio~\cite{virtio}
descriptor rings is the standard answer to MCU/MPU asymmetric
multiprocessing, with NXP's RPMsg-Lite~\cite{rpmsg-lite} as the
vendor-slimmed variant for parts like ours. The comparison is one
of generality versus fit. RPMsg provides dynamic endpoint creation,
variable-length messages, and name-service discovery --- machinery
for systems where the set of services is open. Our transport is a
fixed-function alternative sized to one service: a 20-byte typed
message, two statically laid-out rings whose complete state is
under 1~KB apiece, and a single preallocated exchange slot ---
small enough that the whole shared-memory contract is one header
with compile-time asserts, and simple enough that the
reset-resilience argument (\cref{sec:design:ring}) is inspectable.
We would not argue against RPMsg for a multi-service design; we do
argue that when the service boundary is known and singular,
the fixed layout buys auditability and RAM that generality spends.

Zephyr's own IPC service~\cite{zephyr} with the \texttt{icmsg}
backend is closer kin: also SPSC over shared RAM. Three things
separate this work: our ring carries typed fixed-size slots rather
than a byte-stream packet buffer (no framing layer to verify); our
indices live in the shared region itself, which is what makes the
application core's reboot invisible to the transport; and, more
prosaically, neither Zephyr 3.7's IPC service nor its \texttt{mbox}
driver supports this SoC's second core --- part of the out-of-tree
gap \cref{sec:background:zephyr} documents. Classic dual-OS AMP ---
Linux plus an RTOS on a Cortex-A/Cortex-M pair, the i.MX-class
offload literature --- addresses a different asymmetry (rich OS
versus RTOS); the MCU-to-MCU case, where \emph{both} cores are
resource-poor but unequally so, is comparatively unexamined, and is
exactly where treating the asymmetry as a design input pays.

\subsection{Lock-Free SPSC Queues}
\label{sec:related:spsc}

That a single-producer/single-consumer ring needs no lock is
Lamport's result~\cite{lamport77}; the modern refinements are about
memory systems, not logic. FastForward~\cite{fastforward} slips
cache lines between producer and consumer; the LMAX
Disruptor~\cite{disruptor} pads sequence counters onto private
cache lines --- the same reasoning behind our 64-byte index
separation, applied prophylactically on a non-coherent-cache MCU.
The transfer of this server-class literature down to a 150~MHz
microcontroller is mostly a story of what \emph{disappears}: no
cache coherence protocol to reason about, but also no C11 atomics
library one can assume maps to something sensible across two
masters --- leaving explicit \texttt{DMB} placement per the ARMv8-M
memory model~\cite{armv8m} as the entire ordering story, which our
QEMU-hosted suite cannot validate (\texttt{DMB} is a no-op there)
and the board soak therefore must.

\subsection{Cross-Core Protection on Cortex-M}
\label{sec:related:protection}

MPU-based isolation on Cortex-M is well studied within one core:
MINION~\cite{minion} switches per-task memory views,
ACES~\cite{aces} compiles applications into automatically derived
compartments, and PSA~\cite{psa} standardises isolation levels for
TrustZone-M parts. Phase~3's protection question is the
\emph{cross-core} variant --- one core's MPU guarding another
core's memory --- under an asymmetry those systems do not face: the
core most in need of constraint (the application core, where
arbitrary product code runs) is the one with no MPU at all. Our
contribution here is less a mechanism than an honest scoping of
what the mechanism can mean on real silicon
(\cref{sec:design:mpu}), a statement we have not found made plainly
in vendor documentation for this class of part.

\subsection{Inference Offload Boundaries}
\label{sec:related:offload}

On application processors, inference-as-a-service behind a process
boundary is normal --- Android's NNAPI~\cite{nnapi} routes requests
to a driver process; server-class systems put models behind RPC.
TFLM~\cite{tflm} and CMSIS-NN~\cite{cmsis-nn} on MCUs assume the
caller and the runtime share an address space and a core. Phase~3
occupies the point between: a hardware-enforced (one-directionally,
\cref{sec:design:mpu}) service boundary at MCU scale, where the
``RPC'' is a 20-byte message plus a shared slot and costs
15~\textmu s of transport round-trip against millisecond service
times. The scheduling half of the story --- remote requests
entering the same priority space as local jobs --- extends the
Phase~2 scheduler~\cite{synapticos-p2} across the core boundary
rather than replacing it.

\section{Discussion}
\label{sec:discussion}

\subsection{Case Studies: Two Defects Only the Board Could Find}
\label{sec:discussion:defects}

The Phase~2 paper argued that QEMU's cooperative determinism finds
concurrency bugs emulation is \emph{stronger} at
finding~\cite{synapticos-p2}. Phase~3 supplies the complementary
evidence: two defects that no emulator would ever surface, because
both live in silicon behaviour below the architecture level. We
reconstruct them in detail because each invalidates a design
pattern that looks perfectly reasonable in code review.

\textbf{Case 1: Releasing a core into erased flash wedges the whole
chip.} The original single-core fallback was the obvious design:
release CPU1, wait for the handshake, time out, log, continue.
On this part it is unsurvivable. Erased flash on the MCXN947 reads
as ECC errors at the bus level, and CPU1's very first act after
release is a vector fetch from its (blank) flash bank --- which
stalled the flash system that CPU0 was executing in place from. The
observable symptom was a chip with \emph{no serial output ever}
(deferred logs never flushed) and a dead debug port (CoreSight
returning fault acknowledgements on every access): indistinguishable
from destroyed hardware, recoverable only by ISP-mode reflash. The
diagnosis chain is worth recording --- \texttt{hello\_inference}
booted (so not hardware), the same dual-model image booted when
bank~1 was programmed (so not the image), the debugger's bus-level
fault signature pointed at the interconnect, and the erased-flash
theory completed it. The fix moves the image check \emph{before}
the release and does it through the flash management controller's
command interface (ROM API \texttt{FLASH\_VerifyErase} +
\texttt{FLASH\_Read} + vector sanity), which is safe against erased
pages in a way bus reads are not. The design lesson generalises to
any XIP multi-core part: \emph{never release a core into memory you
have not proven executable, and prove it without bus reads}.

\textbf{Case 2: A Kconfig select silently disarmed the MPU guard.}
The cross-core guard was originally declared where Zephyr wants
static memory attributes declared: a devicetree memory-attribute
node marking CPU1's RAM read-only, compiled into the static MPU
table. It passed its self-test for days --- until the flash driver
was enabled to implement the Case-1 fix, whereupon \texttt{syn mpu
test} failed. The chain: \texttt{SOC\_FLASH\_MCUX} selects
\texttt{MPU\_ALLOW\_FLASH\_WRITE}, which redefines the attribute
macro the devicetree node compiles through
(\texttt{REGION\_FLASH\_ATTR}) from read-only to read-write ---
turning a protection declaration into a no-op, with no warning, in
a different subsystem from the one being edited. The shipped guard
is programmed at runtime into the last MPU region slot with an
explicit read-only access encoding, immune to Kconfig macro
politics. Two lessons: first, on Zephyr, \emph{devicetree-declared
protection is only as strong as every Kconfig select that can
touch its attribute macros}; second --- the one we now treat as
policy --- \emph{a protection mechanism needs a self-test invocable
after every configuration change}, because this regression was
caught only because \texttt{syn mpu test} existed and was run
routinely.

\subsection{What the Asymmetry Bought}
\label{sec:discussion:asymmetry}

It is worth stating what fell out of treating the capability
asymmetry as a design input rather than fighting it. The service
architecture was not merely \emph{compatible} with CPU1's missing
FPU/DSP --- it was \emph{selected} by it, and the result is a
cleaner boundary than a symmetric design would have produced: all
tensor math, model state, and accelerator access live behind one
interface with exactly one implementation. The 64~KB RAM budget
forced zero-copy staging, which removed a copy from the hot path.
The missing TrustZone forced offset-based (never pointer-based)
payload references, which is also what makes the layout
position-independent and assertable. The missing MPU could not be
converted into a benefit --- but scoping it honestly produced a
protection model we can defend, rather than one we would have to
qualify under questioning.

\subsection{Limitations}
\label{sec:discussion:limits}

As in the previous papers, every known gap in one place.

\textbf{Stub NPU baseline.} Every latency involving the model stage
brackets the deterministic stub kernel on real silicon: these are
runtime-and-transport baselines that regression-pin the system's
overhead, not inference throughput claims. The eIQ Neutron invoke
path remains the top integration item. The IPC mechanism numbers
(boot, handshake, ring cost, RTT) are stub-independent.

\textbf{One request in flight.} The exchange slot serialises
cross-core inference; concurrent callers on CPU1 queue on a local
mutex. Ring capacity is not the limit --- the slot is, by design
(\cref{sec:design:infer}). Multi-slot exchange is mechanical future
work if a workload demands it.

\textbf{Handlers share the dispatch thread.} A long inference
delays heartbeat dispatch on CPU0, visibly: it is the measured
81~\textmu s RTT worst case (\cref{sec:eval:rtt}). Moving serving
off the dispatch thread would shrink the tail at the cost of a
thread and a queue; the trade is documented and currently taken in
favour of simplicity.

\textbf{Protection is one-directional and write-only.} CPU1 has no
MPU (silicon); CPU0 reads of CPU1 RAM are not blockable (ARMv8-M
background map). Integrity of CPU1's memory against CPU0 bugs is
enforced; the converse, and confidentiality in either direction,
are not (\cref{sec:design:mpu}).

\textbf{Three acceptance rows are not fully board-demonstrated.}
NPU duty cycle measured $\sim$6\% on a criterion presuming
saturation (\cref{sec:eval:accept}); the error-propagation negative
path is unit-tested but never fired organically on the board; the
staged cross-core priority-preemption experiment was not run.
All three are stated in \cref{tab:accept} at measured strength.

\textbf{CPU1-only reset not independently staged.} Whole-chip
resets exercise the rejoin path (eleven times, clean); a mid-run
reset of CPU1 alone has no button on this board and awaits a
software-triggered experiment.

\subsection{Roadmap}
\label{sec:discussion:roadmap}

Phase~4 turns to model management and over-the-air updates, for
which this phase's flash-controller work is direct groundwork ---
with one recorded hazard: the OTA staging slot and CPU1's flash
bank must not overlap, and the check is on the books before any
OTA write path is enabled. The standing backlog carries the
PowerQuad wrapper optimisation (the Phase~2 miss), camera and LCD
bring-up for an end-to-end vision demo, and the Neutron invoke
path that converts every stub-labeled number in this series into a
silicon claim.

\section{Conclusion}
\label{sec:conclusion}

We presented the Phase~3 dual-core architecture of SynapticOS,
which makes cross-core inference offload an operating-system
service on a commodity dual-core microcontroller whose second core
--- lacking FPU, DSP, TrustZone, and MPU --- is treated as a design
input rather than an obstacle. The AI runtime is confined to the
capable core; the application core reaches it through a
message-based interface with system-call semantics: models resolved
by name, requests carrying the scheduler's priority classes into a
single cross-core priority space, errors and timeouts propagating
to return values, and tensors staged zero-copy in a shared slot
sized by the application core's 64~KB reality. The transport is a
pair of lock-free SPSC rings in shared SRAM --- single-writer
indices, free-running counters, DMB-only ordering, no cross-core
atomics --- whose residence in shared memory makes an
application-core reboot invisible to the conversation.

Measured on the FRDM-MCXN947: 1{,}514~\textmu s application-core
boot and 2{,}554~\textmu s handshake, bit-identical across eleven
consecutive resets; 15~\textmu s typical / 81~\textmu s worst-case
round-trip against a 50~\textmu s budget, with the tail explained
rather than excluded; 25 cycles per ring push; a 1{,}913-serve
two-model soak with zero errors (stub-NPU baselines, labeled); a
fault-injection-verified one-directional MPU guard whose
architectural limits --- no CPU1 MPU, no read blocking under the
ARMv8-M background map --- are stated plainly; and a single-core
fallback that survives a blank second flash bank because it proves
the bank executable through the flash controller before releasing
the core. The dual-core firmware costs 98.9~KB of flash on the
runtime core and 32.4~KB on the application core; the test suite
grows to 108 cases in 13 suites at 100\% pass. Two board-found
defects --- a whole-chip wedge from releasing into erased flash,
and a Kconfig select that silently disarmed a devicetree-declared
MPU guard --- are reconstructed in full as the phase's most
transferable results.

SynapticOS v0.3.0, the out-of-tree CPU1 board port, the test
suite, the QEMU and FRDM measurement artifacts (including the raw
serial transcripts behind every board number in this paper), and
the LaTeX sources of this paper are released under Apache~2.0 at
\url{https://github.com/Dimitrios-Kafetzis/SynapticOS}.

\bibliographystyle{IEEEtran}
\begingroup
\emergencystretch=1em
\bibliography{refs}

\begin{thebibliography}{10}
\providecommand{\url}[1]{#1}
\csname url@samestyle\endcsname
\providecommand{\newblock}{\relax}
\providecommand{\bibinfo}[2]{#2}
\providecommand{\BIBentrySTDinterwordspacing}{\spaceskip=0pt\relax}
\providecommand{\BIBentryALTinterwordstretchfactor}{4}
\providecommand{\BIBentryALTinterwordspacing}{\spaceskip=\fontdimen2\font plus
\BIBentryALTinterwordstretchfactor\fontdimen3\font minus
  \fontdimen4\font\relax}
\providecommand{\BIBforeignlanguage}[2]{{%
\expandafter\ifx\csname l@#1\endcsname\relax
\typeout{** WARNING: IEEEtran.bst: No hyphenation pattern has been}%
\typeout{** loaded for the language `#1'. Using the pattern for}%
\typeout{** the default language instead.}%
\else
\language=\csname l@#1\endcsname
\fi
#2}}
\providecommand{\BIBdecl}{\relax}
\BIBdecl

\bibitem{synapticos-p1}
D.~Kafetzis, ``{SynapticOS}: An inference-first runtime architecture for neural
  processing units on resource-constrained microcontrollers,'' Preprint,
  SynapticOS Project. \url{https://github.com/Dimitrios-Kafetzis/SynapticOS},
  2026, phase~1 paper; LaTeX sources and measurement artifacts in the
  repository.

\bibitem{synapticos-p2}
------, ``Inference pipelines as operating-system objects: Priority scheduling
  and constant-footprint streaming for microcontroller neural inference,''
  Preprint, SynapticOS Project.
  \url{https://github.com/Dimitrios-Kafetzis/SynapticOS}, 2026, phase~2 paper;
  LaTeX sources and measurement artifacts in the repository.

\bibitem{nxp-mcxn947}
{NXP Semiconductors}, ``{MCX} {N}947 reference manual,'' Document MCXNX4XRM,
  Rev.~5, 2024.

\bibitem{zephyr}
{Zephyr Project}, ``Zephyr {RTOS},'' \url{https://zephyrproject.org}, 2024,
  version 3.7.0 LTS.

\bibitem{lamport77}
L.~Lamport, ``Concurrent reading and writing,'' \emph{Communications of the
  ACM}, vol.~20, no.~11, pp. 806--811, 1977.

\bibitem{armv8m}
{Arm}, ``Armv8-{M} architecture reference manual,'' Document DDI0553B.y, 2024.

\bibitem{fastforward}
J.~Giacomoni, T.~Moseley, and M.~Vachharajani, ``{FastForward} for efficient
  pipeline parallelism: A cache-optimized concurrent lock-free queue,'' in
  \emph{ACM SIGPLAN Symposium on Principles and Practice of Parallel
  Programming (PPoPP)}, 2008.

\bibitem{disruptor}
M.~Thompson, D.~Farley, M.~Barker, P.~Gee, and A.~Stewart, ``Disruptor: High
  performance alternative to bounded queues for exchanging data between
  concurrent threads,'' LMAX Exchange technical paper, 2011.

\bibitem{openamp}
{OpenAMP Project}, ``{OpenAMP}: Open asymmetric multi-processing framework,''
  \url{https://www.openampproject.org}, 2024.

\bibitem{virtio}
R.~Russell, ``virtio: Towards a de-facto standard for virtual {I/O} devices,''
  \emph{ACM SIGOPS Operating Systems Review}, vol.~42, no.~5, pp. 95--103,
  2008.

\bibitem{rpmsg-lite}
{NXP Semiconductors}, ``{RPMsg-Lite}: Lightweight remote processor messaging,''
  \url{https://github.com/nxp-mcuxpresso/rpmsg-lite}, 2024.

\bibitem{minion}
C.~H. Kim, T.~Kim, H.~Choi, Z.~Gu, B.~Lee, X.~Zhang, and D.~Xu, ``Securing
  real-time microcontroller systems through customized memory view switching,''
  in \emph{Network and Distributed System Security Symposium (NDSS)}, 2018.

\bibitem{aces}
A.~A. Clements, N.~S. Almakhdhub, S.~Bagchi, and M.~Payer, ``{ACES}: Automatic
  compartments for embedded systems,'' in \emph{USENIX Security Symposium},
  2018.

\bibitem{psa}
{Arm}, ``Platform security architecture: Security model and isolation levels,''
  \url{https://www.psacertified.org}, 2024.

\bibitem{nnapi}
{Google}, ``Android neural networks {API},''
  \url{https://developer.android.com/ndk/guides/neuralnetworks}, 2024.

\bibitem{tflm}
R.~David, J.~Duke, A.~Jain, V.~Janapa~Reddi, N.~Jeffries, J.~Li, N.~Kreeger,
  I.~Nappier, M.~Natraj, S.~Regev, R.~Rhodes, T.~Wang, and P.~Warden,
  ``{TensorFlow Lite Micro}: Embedded machine learning on {TinyML} systems,''
  in \emph{Proceedings of Machine Learning and Systems (MLSys)}, 2021.

\bibitem{cmsis-nn}
{Arm}, ``{CMSIS-NN}: Efficient neural network kernels for {A}rm {C}ortex-{M}
  cpus,'' \url{https://github.com/ARM-software/CMSIS-NN}, 2023.

\end{thebibliography}
\endgroup

\end{document}